\documentclass{cernrep} 
\usepackage{texnames}
\usepackage[T1]{fontenc}
\begin{document}
\title{Model Inference with Reference Priors}
 
\author{M.~Pierini$^{1}$, H.~Prosper$^{2}$, S.~Sekmen$^{2}$, M.~Spiropulu$^{1,3}$}
\institute{(1) CERN, Geneva, Switzerland\\
(2) Florida State University, Tallahassee (FL), USA \\
(3) Caltech, Pasadena (CA), USA}

\maketitle % this produces the title block

\begin{abstract}
We describe the application of model inference based on reference
priors to two concrete examples in high energy physics: the
determination of the CKM matrix parameters $\bar \rho$ and $\bar
\eta$ and the determination of the parameters $m_0$ and $m_{1/2}$
in a simplified version of the CMSSM SUSY model. We show how a
1-dimensional reference posterior can be mapped to the $n$-dimensional ($n$-D)
parameter space of the 
given class of models, under a minimal set of conditions on the 
$n$-D function. 
This reference-based function can be used as a prior for the next
iteration of inference, using Bayes' theorem recursively.
\end{abstract}
 
\section{Introduction}

It is typical in high energy physics (HEP) to deal with classes of
models, e.g. new physics extensions of the Standard Model (SM), 
differing by the values of a set of (typically continuous) unknown parameters. 

Given a set of experimental measurements, one would like to 
define the region of the model parameter space that is in
agreement with the data. This is what we refer to as {\it Model
  Inference}. The following ingredients are needed:
\begin{itemize}
\item a theoretical tool that predicts the expected values of
  the measured observables, given a point in the model parameter space;
\item a multi-dimensional likelihood, built from the available
  measurements;
\item and a statistical procedure that evaluates the level of agreement
  between the data and the predictions.
\end{itemize}

While the first and second steps are not controversial, the third
step is often polemical and is subject to some degree of arbitrariness.
Two main approaches are typically followed: Bayesian, which computes
the posterior probability of the expected values of
the model parameters given the likelihood and a prior probability, and 
frequentist, which provides probability statements about possible values of
the {\it measurements} given the observed data and assumed values
of the model parameters. 

Historically, most high energy physicists have preferred frequentist statistics because (they
say) it allows one to extract statistical information from data without the
need for  subjective input. In this sense, these physicists are victim of the utopian idea
of an analyst-free analysis, in which the  ``data speak for themselves", 
independently of the personal opinion and
judgement 
of the physicists who perform the analysis. However, we are rudely awakened from this 
utopian dream 
on a daily basis as anybody who has had to evaluate a 
systematic uncertainty can confirm~\footnote{About 10\% of the
 {\tt hep-ex} papers on {\tt INSPIRE} match the search for the word {\it
   assume}, which is quite far from the analyst-free paradigm of
   our dreams.}.
Beyond this simple fact, we
also tend to underestimate how strongly the subjective beliefs of the
analyst enters the earlier stages of an analysis, as for instance when
we {\it define} the form of the likelihood. Physicists quote results as 
$m \pm \sigma$, where $m$ and $\sigma$ summarize the result of,
perhaps, a likelihood-ratio-based analysis, which already implies assumptions
about the form of the likelihood~\footnote{The use of the likelihood
ratio to quote the range $[m-\sigma, m+\sigma]$ for the unknown $x$
as the 68\% confidence region implies that we can define a one-to-one
transformation $x \rightarrow f(x)$ such that the likelihood is
Gaussian in $f(x)$.}. When estimating the systematic uncertainty, we typically sum the 
different contributions in quadrature, implying that the systematic errors are
uncorrelated and, more importantly, that they may be treated as if they are statistical. This
may be true of systematic uncertainties that arise ultimately from other statistics; but many
systematic uncertainties are ``assigned" based on judgement or official policy.

We push this even further when we perform phenomenological analyses. While
connecting the parameters of a model to the experimental observables,
we often need to know a set of additional quantities (theoretical
nuisance parameters) which are not measurable, but which may be known with 
some uncertainty through a theoretical calculation. This is 
the case, for instance, for the non-perturbative QCD parameters determined using
lattice QCD calculations. In order to take into account the uncertainty
on the predictions correctly, a Bayesian analyst would introduce a prior
probability density function (pdf) for the theoretical nuisance
parameters based on the best judgement of the theorist. While this is considered 
{\it dangerously subjective} in by many high energy physicists, the same physicists consider it safe to modify the likelihood to take account of the theoretical uncertainty on predictions. 
This breaks the
objective-frequentist-physicists paradigm twice: i) the functional
form used to account for theoretical uncertainty is no less subjective than 
the prior of a Bayesian analysis and ii)
the likelihood loses its deep and precise meaning of that function obtained by inserting the
observations 
 into the probability density function describing possible observations. Nobody ever did (and it is likely that nobody ever will) measure the 
theoretical nuisance parameters --- indeed, many such parameters such as
the factorization and renormalization scales are pure artifacts of our current reliance on 
perturbation theory in theoretical calculations. 
As a matter of fact, a physicist
performing data analysis is forced to make assumptions. And there is
nothing wrong with that as long as the assumptions are clearly stated. 
The problems come when the assumptions are hidden in the procedure
and not transparent to the people not directly involved in the
analysis. 

The contrasting attitudes described above can be summarized 
in terms of the following two perceived problems:
\begin{itemize}
\item For some high energy physicists, introducing a prior
is unacceptable because it brings subjectivity into science. 
{\it ``The origin of the problem lies in the very first Bayesian
  assumption, namely that unknown model parameters are to be 
understood as mathematical objects distributed according to PDFs,
which are assumed to be known: the priors. Obviously, the choice of
the priors cannot be irrelevant; hence, the Bayesian treatment is
doomed to lead to results which depend on the decisions made, 
necessarily on an unscientific basis, by the authors of a given analysis, 
for the choice of these extraordinary PDFs.'' }~\cite{Charles}.

\item For some statisticians, a meaningful statistical analysis is not 
possible in the absence of an analysis procedure that allows one to 
incorporate a priori knowledge in a coherent way. {\it ``The frequentist approach to
  hypothesis testing does not permit researchers to place
  probabilities of being correct on the competing hypotheses. This 
is because of the limitations on mathematical probabilities used by 
frequentists. For the frequentists, probabilities can only be defined 
for random variables, and hypotheses are not variables (they are 
not observables)... This limitation for frequentists is a real
drawback because the applied researcher would really like to be 
able to place a degree of belief  on the hypothesis. He or she would 
like to see how the weight of evidence modifies his/her degree of 
belief (probability) on the hypothesis being true.''}~\cite{Press}.
\end{itemize}

The use of reference priors~\cite{refPriors} is emerging  as
a concrete way to solve the two problems. While a detailed discussion of
the reference priors is beyond the scope of this paper, we highlight
here their most appealing properties.  

The main concern against the use of a Bayesian analysis in HEP 
is related to a priori ignorance, more than a priori knowledge. 
Whenever a priori knowledge is available 
(e.g. the measurement of the luminosity, which is used to translate an observed
signal yield into a cross section measurement), there is a
general consensus that an \emph{evidence-based} prior should be 
used. The real issue is how we should parameterize 
"ignorance".  The use of  a flat prior, a HEP standard, is not quite the right answer.
Reference priors can be seen as a model of ignorance in the sense that, on average, they
maximize the influence of the likelihood relative to the prior; hence they are a
solution to this problem. More precisely, for a given likelihood, the reference prior is the
prior function that {\it on average} maximizes the asymptotic Kullback-Leibler
divergence~\cite{KL} between the prior and the posterior, hence enhancing the
role of the likelihood (the data) over the prior.  This is exactly the
kind of behavior that we would like for a model of ignorance.
And this is what we assume the flat prior does for us, when we use it.
Unfortunately, the flatness of the prior is not invariant
under reparameterization. Unlike the flat prior, reference
priors give reparameterization-invariant results in the cases
typically considered in HEP (e.g. one-to-one transformations for which
the Jacobian is not singular~\cite{RepInv}). The use of 
reference priors in HEP has been recently proposed in Ref.~\cite{LucHarrison},
where the application in the case of a counting experiment
is discussed. This has been applied to real LHC data, in one of
the CMS Supersymmetry (SUSY) searches~\cite{Razor}.

In the following, we apply the procedure described in
Ref.~\cite{LucHarrison} to two specific cases: i) the determination
of the parameters $\bar \rho$ and $\bar\eta$ (at fixed $A$ and $\lambda$) of a {\it simplified} CKM 
matrix 
and ii) the determination of  the parameters in 
the case of a SUSY model~\footnote{For simplicity, we take a
  simplified CMSSM with $m_0$ and $m_{1/2}$, fixing $A_0=0$,
  $\tan\beta=10$, and positive $\mu$.}. In both cases, as
an illustration, we limit
the discussion to the determination of two parameters. The generalization to $n>2$ dimensions is computationally
more demanding, but conceptually equivalent. 
In both cases, we start from one experimental measurement, for which
the likelihood can be analytically modeled without too much
arbitrariness. We briefly describe the derivation of the reference
posterior, following Ref.~\cite{LucHarrison}. We then map the
1-D posterior into a $n$-D ($n=2$ in our examples) function of the model
parameters, introducing the {\it look-alike} (LL) prescription. This
function, based on a reference prior, can then be used  as the 
prior in a recursive application of Bayes' theorem to include
other measurements. 

\section{The reference posterior for a 1-D analysis}

When looking for a signal, produced by the process under study,
we are confronted with a Poisson count of a signal on top
of a background coming from other physics processes. The likelihood
for the signal, in the absence of a background,  is described by a
Poisson function. In the presence of a background the  likelihood 
asymptotically converges to a Gaussian density. Under these 
conditions, the reference prior is Jeffrey's prior for a 
Poisson likelihood, $\pi(\theta)\sim 1/\sqrt{\theta}$.

This is the case for the exclusive measurement of $V_{ub}$ from 
$B \to \pi \ell \nu$ decays. What one measures is the branching ratio
$BR(B \to \pi \ell \nu)$, which is the related to the the absolute value of 
the CKM matrix element $V_{ub}$ as:
\begin{equation}
|V_{ub}|^2 = \frac{BR(B \to \pi \ell \nu)}{\Gamma_B F(B \to \pi)},
\end{equation}
where evidence-based priors are available both for the
width of the $B$ meson $\Gamma_B$ (from other measurements) and
the $B \to \pi$ form factor $F(B \to \pi)$ (from theory). 
One can determine the reference posterior for the $BR$
using $\pi(BR)\sim 1/\sqrt{BR}$.

For SUSY searches, 
one looks for a signal yield $s$ in a signal-sensitive {\it box}, 
defined by a selection using signal-vs-background separating 
variables. One observes a yield $n = s+\mu$, where $\mu$ is the
background surviving the signal-enhancing selection. The expected
background $\bar \mu$ is estimated from a sideband region where
no signal is expected, where the observed yield in the sideband is $y$ and the scaling
factor $b$ is such that $b \mu$ is the expected background yield in the sideband.
In formulas:
\begin{itemize}
\item the likelihood is $p(n|s,\mu) = (s+\mu)^n e^{-(s+\mu)} / n!$,
\item the prior for $\mu$ is $\pi(\mu)=b(b\mu)^{y-1/2}e^{-b\mu} / \Gamma(y+1/2)$ and
\item the prior on $s$ is $\pi(s) = \pi(s|\mu) \propto 1/\sqrt{s+\mu}$,
\end{itemize} 
where $\Gamma(x)$ is the gamma function. 

Once the 1-D reference-posterior is derived as described above, we
translate this into an $n$-D function of the model parameters. 
While rigorous algorithms exist to build the $n$-D reference
prior~\cite{refPriors}, we follow here a computationally simpler {\it heuristic}
construction, described below, which we call the {\it
  look-alike prescription}. 

 \section{Look-alike prescription}

Mapping the 1-D reference posterior to the $n$-D space of the model
parameter under consideration could be achieved by demanding
that  the $n$-D 
pdf satisfy two requirements:
\begin{itemize}
\item all models predicting the same values for the parameter
  $x$ ($V_{ub}$ and
  $s$ in our two examples) associated with the posterior density $P(x)$ are equi-probable and
\item the $n$-D function should be such that it maps back to a 1-D function 
  $P'(x)$ identical to the $P(x)$ with which we started. 
  \end{itemize}
  Given the mapping $\theta: \rightarrow x$ predicted by the physics model, these requirements are sufficient to map $P(x)$ to $\pi(\theta)$.
We first 
    write the $n$-D function as $\pi(\theta) = K(x(\theta)) \times
  P(x(\theta))$.
The computation of $K(x(\theta))$ goes as follows:
\begin{eqnarray}
P'(\tilde{x}) &=& \int d\theta \, P(x(\theta)) \, K(x(\theta)) \,
\delta(\tilde{x}-x(\theta)), \nonumber \\
&=& P(\tilde{x}) K(\tilde{x}) \int d\theta \, \delta(\tilde{x}-x(\theta))
= P(\tilde{x}),
\end{eqnarray}
where the last equality follows from the second condition. This
implies that
\begin{equation}
K(\theta) = \frac{1}{ \int d\theta \, \delta(\tilde{x}-x(\theta))},
\end{equation} 
which is the surface of the region spanned by the look-alike  (LL) models,
that is, models giving the
same value $\tilde{x}$~\footnote{The challenge of
  generalizing this approach to a generic $n$-D problem is the calculation
  of this surface term.}.

\begin{figure}
\centering\includegraphics[width=.45\linewidth]{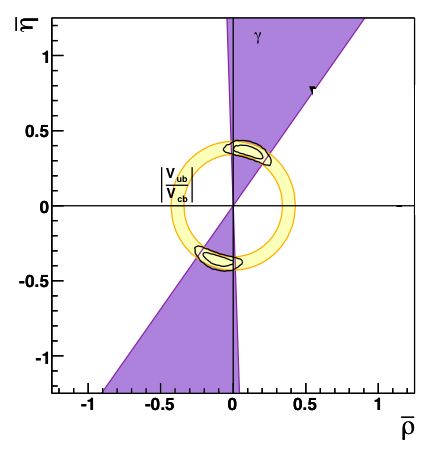}
\centering\includegraphics[width=.45\linewidth]{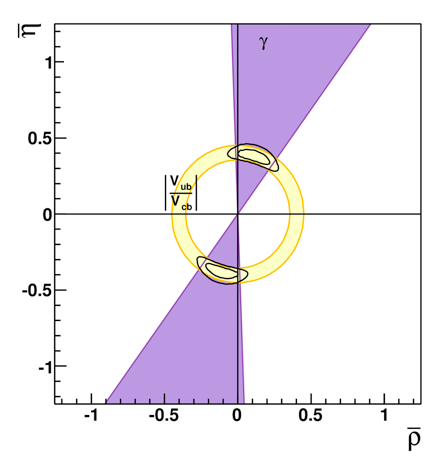}
\caption{Result for the 2-D allowed region for the CKM parameters $\bar
  \rho$ and $\bar \eta$, obtained using the reference posterior for
  $V_{ub}$ and the LL prescription (left), or using flat priors for $\bar
  \rho$ and $\bar \eta$ (right).}
\label{fig:UTres}
\end{figure}

The case of $V_{ub}$ is useful because it allows us to explain how this works in practice.
All the models such that $\bar \rho^2 + \bar \eta^2 =k$ predict 
the same value of $|V_{ub}|$. This makes them LL models, by our
definition.  The LL domain
is a circle centered at $0$ with radius $k$. The $n$-D function is therefore,
\begin{equation}
\pi(\bar \rho, \bar \eta) = \frac{ P(V_{ub}(\bar \rho^2 + \bar
  \eta^2))} {2\pi \sqrt{\bar \rho^2 + \bar \eta^2}},
\label{eq:LLcontour}
\end{equation}
where $P(V_{ub})$ is the reference posterior for $|V_{ub}|$. 
The function $\pi(\bar \rho, \bar \eta)$ is then used as the prior to fit the CKM
matrix~\cite{UTFIT} including the measurement of the CKM phase
$\gamma$. This step gives the allowed 
region for $\bar \rho$ and $\bar \eta$ shown in the left plot of
Fig.~\ref{fig:UTres}),  which is to be compared to a similar plot obtained
using flat priors for $\bar \rho$ and $\bar \eta$ (right plot of
Fig.~\ref{fig:UTres}). 
The results of these two calculations are consistent.  However,
the reference posterior for  $|V_{ub}|$ provides a
 more solid foundation for determining the prior to associate with the CKM
 parameters. 
% allowing to specify what one is ignorant about (the
% measured value of $|V_{ub}|$, while than generically the CKM matrix itself).

\begin{figure}
\centering\includegraphics[width=.45\linewidth]{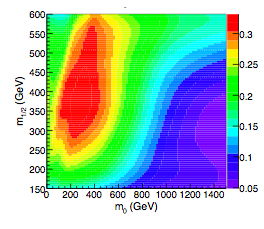}
\centering\includegraphics[width=.45\linewidth]{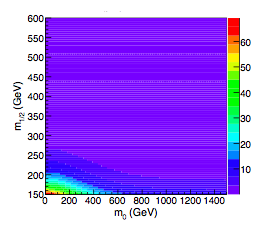}
\caption{$\epsilon(m_0, m_{1/2})$ and $\sigma(m_0, m_{1/2})$ 
functions in the case of a hypothetical SUSY  search~\cite{ourPaper}.}
\label{fig:mSugraLL}
\end{figure}

Our second case study, which uses a simplified version of the mSUGRA model,
 is more complicated since
Eq.~\ref{eq:LLcontour} cannot be solved analytically in the case of a
generic search for new physics. In this case, the LL domain is given by all
the models predicting the same expected signal yield $s$.
The expected signal yield as a function of the model parameters can be written as 
$s(m_0, m_{1/2}) = \epsilon(m_0, m_{1/2}) \sigma(m_0, m_{1/2})
\cal{L}$, where only the luminosity $\cal{L}$ is a constant, while
both the cross section $\sigma$ and the efficiency of the applied
selection $\epsilon$ depends on the
features of the model (e.g. the masses of the SUSY
particles), and hence on the model parameters. The function $\sigma(m_0,
m_{1/2})$ can be computed from the SUSY Lagrangian, while
$\epsilon(m_0, m_{1/2})$ has a non-trivial dependence on the models,
through several effects connected to the detector response. For
instance, a model with large (small) mass differences would give 
a large (small) value of $\epsilon$, since harder (softer) spectra 
for the visible particles produced in the SUSY decay chain will
have larger (smaller) chance to survive the kinematic cuts. In general, the connection
between the features of the model and the detector performance
produces non-analytical iso-yield contours for the LL domains. This is
illustrated in Fig.~\ref{fig:mSugraLL}, where $\epsilon(m_0, m_{1/2})$ and
$\sigma(m_0, m_{1/2})$ are shown in the case of a hypothetical SUSY  
search~\cite{ourPaper}.

On the other hand, all the iso-yield contours have infinite length,
resulting in constant $K(m_0, m_{1/2})$ if one considers the full domain
for $m_0$ and $m_{1/2}$, and approximately constant if one uses a
large-enough domain in practice~\footnote{In case the measurement
  points to particular region of the plane, i.e. when there is hint of a
  signal, one could use the Savage prescription and cut the
plot where the likelihood drops to negligible values. In absence of a
signal hint, the situation is complicated by the fact that the
likelihood peaks at infinite values of $m_0$ and $m_{1/2}$, where the
SUSY particles are so heavy that they decouple from the SM ones,
effectively recovering the SM limit.}. We can then take $K(m_0,
m_{1/2})$ as a constant and show how the method works. It has to
be clearly stated that this is an approximation, and that the computation of
the surface term of Eq.~\ref{eq:LLcontour} is the main challenge in the
applicability of the proposed method in its exact form (see Ref.~\cite{ourPaper} for details).

\begin{figure}
\centering\includegraphics[width=.9\linewidth]{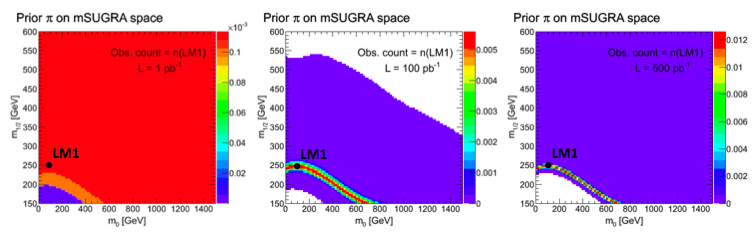}
\caption{Result for the 2-D function mapped from the 1-D reference
  posterior in the case of 1 pb$^{-1}$ (left), 100 pb$^{-1}$ (center), and
  500 pb$^{-1}$ (right) integrated luminosity. We assume a {\it
    measurement} perfectly in agreement  with the expectation from the 
    true model, corresponding to $m_0=60$, $m_{1/2}=250$.}
\label{fig:mSugraResult}
\end{figure}

For illustration, we take the CMS low mass (LM) point~\cite{CMSptdr} ($m_0=60$,
$m_{1/2}=250)$ as the {\it  true} state of nature 
and we simulate the case of an experiment giving a result
exactly at the expectation, for low (1 pb$^{-1}$), moderate (100
pb$^{-1}$), and large  (500 pb$^{-1}$) statistics. Figure~\ref{fig:mSugraResult}
shows the 2-D function obtained by the LL prescription. With increasing
sample size, the function shows a peak corresponding to the  {\it true} 
value and to all its (degenerate) LL models, showing the consistency of
the procedure.

\section{Conclusions}

We described the use of the reference prior to 1-D cases (typical of a
HEP measurement) and how this can be used to define an $n$-D function
of the model, induced by the 1-D reference posterior, which may then be
used as a prior for further applications (e.g. to fit to model
parameters). The connection between the 1-D posterior on a measurable
quantity $s$ (e.g. a signal yield on top of a background $b$) and an
$n$-D function of a set of interesting parameters (e.g. the parameters
of a SUSY model) is established through the look-alike prescription, which defines a heuristic
procedure on the basis of two minimal conditions: i) the models
predicting the same expected value for the interesting variable $s$ are
equi-probable and ii) the $n$-D function should map back to the 1-D reference
posterior for $s$, from which we started. This requires the calculation of a surface
term (see Eq.~\ref{eq:LLcontour}), which can be performed
numerically~\cite{ourPaper}. While in specific cases this choice of a
prior might be in conflict with a subjective assessment that could
favor one region of the parameter space over another, it should be stressed that this
{\it Bayesian} approach is likely to give the best frequentist performance because
of the good frequentist properties of reference priors.

We provided two simplified 2-D examples to illustrate
the method, for which computational complications are absent
or marginal. Work is in progress to extend this procedure to more
realistic cases~\cite{ourPaper}. 

\section{Acknowledgment}

The authors would like to thank J.~Berger for the helpful suggestions
and the interesting discussion.

\end{document}